\documentclass{pspum-l}

\usepackage{amssymb,latexsym}
\usepackage{amsmath}
\usepackage{graphicx}
\usepackage{graphics}
\usepackage{extarrows}
\usepackage{float}
\usepackage{cite}
\usepackage{epsfig}
\usepackage{subfig}
\usepackage{mathrsfs}
\usepackage{tensor}

\newcommand{\al}[1]{\begin{align}#1\end{align}}
\newcommand{\bs}{\begin{split}}
\newcommand{\es}{\end{split}}
\newcommand{\eqr}{\eqref}
\newcommand{\mc}{\mathcal}
\newcommand{\pa}[1]{\left(#1\right)}
\newcommand{\pb}[1]{\left[#1\right]}

\numberwithin{equation}{section}

\begin{document}


\title{Chern-Simons Splitting of 2+1D Gauge Theories}
\author{Tuna Yildirim}
\address{Physics and Astronomy Department, The University of Iowa, Iowa City, Iowa 52242}
\curraddr{Department of Physics,
Arizona State University, Tempe, Arizona 85287}
\email{tuna.yildirim@asu.edu}
\thanks{The author was supported in part by NSF Grant \#1067889.}

\date{January 1, 2015}


\begin{abstract}
Geometric quantization of topologically massive and pure Yang-Mills theories is studied in 2+1 dimensions. Analogous to the topologically massive AdS gravity model, both topologically massive Yang-Mills and pure Yang-Mills theories are shown to exhibit a Chern-Simons splitting behavior at large distances. It is also shown that, this large distance topological behavior can be further used to incorporate link invariants.
\end{abstract}

\maketitle


\section{Introduction}

In 2+1 dimensions, Yang-Mills(YM) theory is known to have a mass gap, which makes the theory trivial at very large distances. However, this is not the case for topologically massive Yang-Mills theory(TMYM). In TMYM theory, as Yang-Mills contribution decays exponentially at large distances, the Chern-Simons(CS) term becomes dominant, which leads to a topological theory. For a mass gap of $m$, the scale is given in comparison to $1/m$. This work focuses on the large but finite distance behavior of TMYM and pure YM theories by taking the first order contributions in the $1/m$ expansion into account, neglecting all higher order terms. At this limited scale, both theories have interesting topological behaviors, very similar to the topologically massive AdS-gravity model at corresponding limits.


\subsection{Topologically Massive AdS Gravity}\label{sec:TMG}

For a dynamical metric $\gamma_{\mu\nu}$, the action for the topologically massive AdS gravity model is given by\cite{Deser1982372,Deser2}
\al{
\label{eq:tmgS}
S=\int d^3x \left[ -\sqrt{-\gamma}(R-2\Lambda)+\frac{1}{2\mu}\epsilon^{\mu\nu\rho}\left(  \Gamma^\alpha_{\mu\beta}\partial_\nu \Gamma^\beta_{\rho\alpha} +\frac{2}{3}\Gamma^\alpha_{\mu\gamma} \Gamma^\gamma_{\nu\beta} \Gamma^\beta_{\rho\alpha}   \right)  \right].
}
This action naturally splits into two CS terms\cite{achucarro1986chern,witten1988,Carlip2008272, Carlip2}, by defining
\al{
A^{\pm}{}_{\mu}{}^a{}_b[e]=\omega_{\mu}{}^a{}_b[e] \pm \epsilon^a{}_{bc} e_\mu{}^c,
}
where $e_\mu{}^a$ is the dreibein and $\omega_{\mu}{}^a{}_b[e]$ is the torsion-free spin connection. Then, the action \eqref{eq:tmgS} can be written as
\al{
\label{eq:tmg2CS}
S[e]=-\frac{1}{2}\pa{1-\frac{1}{\mu}}S_{CS}\big[A^+[e]\big]+\frac{1}{2}\pa{1+\frac{1}{\mu}}S_{CS}\big[A^-[e]\big]
}
where 
\al{
S_{CS}[A]=\frac{1}{2}\int \epsilon^{\mu\nu\rho}\pa{A_\mu{}^a{}_b \partial_\nu A_\rho{}^b{}_a + \frac{2}{3} A_\mu{}^a{}_c A_\nu{}^c{}_b A_\rho{}^b{}_a}.
}
For our interests, the main difference between this gravity model and TMYM theory is that the latter has a mass gap, therefore it has a topological behavior only at large distances. But the gravity model is given by CS theory irrespective of the value of $\mu$. In TMYM theory, large distances are obtained by taking large values of $m$. Large $m$, in the sense of near CS limit of TMYM theory, corresponds to small values of $\mu$ in the gravitational analogue. On the other hand, the $\mu\rightarrow\infty$ limit(pure Einstein-Hilbert limit) of the gravity model is analogous to pure YM theory. 

Now, let us focus on the two important limits of \eqref{eq:tmg2CS}. For small values of $\mu$, \eqref{eq:tmg2CS} can be written as a \emph{sum} of two half CS theories as
\al{
\label{eq:tmgcs+cssplitting}
S[e]\approx\frac{1}{2\mu}S_{CS}\big[A^+[e]\big]+\frac{1}{2\mu}S_{CS}\big[A^-[e]\big].
}
In the $\mu\rightarrow\infty$ limit, it is equal to the \emph{difference} between two half CS theories, as
\al{
\label{eq:YMlimit}
S[e]=\frac{1}{2}S_{CS}\big[A^-[e]\big]-\frac{1}{2}S_{CS}\big[A^+[e]\big].
}
The main goal of this work is to investigate whether or not a similar CS+CS type splitting appears in TMYM theory and a CS--CS type splitting in pure YM theory at large distances.


\subsection{Holomorphic Quantization of Chern-Simons Theory}

Before we tackle pure and topologically massive Yang-Mills theories, it would be beneficial to review the geometric quantization of pure Chern-Simons theory. This section will be a quick review, following refs. \citen{Bos:1989kn,Nair:2005iw}.

In this section and all following sections, we will choose the temporal gauge $A_0=0$ for all of the theories we study. We will also use the complex coordinates, where $A_z=\frac{1}{2}(A_1+iA_2)$ and $A_{\bar{z}}=\frac{1}{2}(A_1-iA_2)$. In these coordinates, the conjugate momenta of CS theory are given by
\al{
\Pi^{a z}=\frac{ik}{4\pi} A^a_{\bar{z}}~\ \text{and}~\ \Pi^{a \bar{z}}=-\frac{ik}{4\pi}  A^a_z.
}
Now, we can write the symplectic two-form for pure CS theory as
\al{
\label{eq:omegacs}
\Omega=\frac{ik}{2\pi}\int\limits_\Sigma \delta A^a_{\bar{z}}\delta A^a_z
}
where $\Sigma$ is an orientable two-dimensional surface.

\subsubsection{The Wave-Functional}

We start with choosing the holomorphic polarization that gives $\Phi[A_z,A_{\bar{z}}]=e^{-\frac{1}{2}K}\psi[A_{\bar{z}}]$, where $K$ is the K\"ahler potential, $\Phi$ and $\psi$ are pre-quantum and quantum wave-functionals. K\"ahler potential is given by $K=\frac{k}{2\pi}\int\limits_\Sigma A^a_{\bar{z}} A^a_z$.

Upon quantization we can write
\al{
A_z \psi[A_{\bar{z}}] = \frac{2\pi}{k}\frac{\delta}{\delta A_{\bar{z}}}\psi[A_{\bar{z}}].
}

For the gauge theories that we are interested in, the wave-functional can be factorized as $\psi=\phi\chi$, where $\phi$ is the part that satisfies Gauss' law of the theory and $\chi$ is the gauge invariant part. To find $\phi$, one makes an infinitesimal gauge transformation, then forces the Gauss law constraint. This leads to a condition that is usually solved by some WZW action. Then, the Schr\"odinger's equation can be solved to find $\chi$. The Hamiltonian of pure CS theory in the temporal gauge is zero, hence Schr\"odinger's equation is trivially satisfied. This makes $\chi=1$ a sufficient solution. For the theories that are not scale invariant, $\chi$ is where the scale dependence would be hidden. 

For pure CS theory, the generator of infinitesimal gauge transformations is $G^a=\frac{ik}{2\pi}F^a_{z\bar{z}}$ and the Gauss law is given by $G^a\psi=0$.

Before we continue, we shall parametrize the gauge fields, using Karabali-Nair\cite{Karabali:1996iu} parametrization, as
\al{
\label{eq:para}
A_z=U^{\dagger-1}\partial_z U^\dagger ~\text{and}~ A_{\bar{z}}=-\partial_{\bar{z}}UU^{-1}.
}
Here $U \in SL(N,\mathbb{C})$ and it gauge transforms as $U\rightarrow gU$, where $g\in \mc{G}$ and $\mc{G}$ is the gauge group. $U$ can be written as
\al{
\label{eq:U}
U(x,0,C)=\mc{P}exp\pa{-\underset{C}{\ \ \int_0^x}(A_{\bar{z}}d\bar{z}+\mc{A}_zdz)},
}
where $\mc{A}_z$ satisfies $\partial_z A_{\bar{z}}-\partial_{\bar{z}}\mc{A}_z+[\mc{A}_z,A_{\bar{z}}]=0$. This flatness condition is what makes \eqref{eq:para} a good parametrization, since it makes $U$ invariant under small deformations of the path $C$ on $\Sigma$.  From \eqref{eq:U}, it follows that
\al{
\label{eq:scriptA}
\mc{A}_z=-\partial_z U U^{-1} ~\text{and}~ \mc{A}_{\bar{z}}= U^{\dagger-1}\partial_{\bar{z}} U^{\dagger}.
}

Now, we make an infinitesimal gauge transformation on $\psi$, as
\al{
\bs
\delta_\epsilon \psi[A_{\bar{z}}]=&\int d^2z\ \delta_\epsilon A^a_{\bar{z}}\ \frac{\delta\psi}{\delta A^a_{\bar{z}}} \\
=& \int d^2z\ \epsilon^a \pa{ \partial_{\bar{z}} \frac{\delta}{\delta A^a_{\bar{z}}} + i f^{abc} A^b_{\bar{z}}  \frac{\delta}{\delta A^c_{\bar{z}}} }\psi\\
=& -\frac{k}{2\pi} \int d^2z\ \epsilon^a (F^a_{z\bar{z}} - \partial_z A^a_{\bar{z}}) \psi.
\es
}
After applying the Gauss law, one gets
\al{
\label{eq:infg}
\delta_\epsilon \psi[A_{\bar{z}}]= \frac{k}{2\pi} \int d^2z\ \epsilon^a  \partial_z A^a_{\bar{z}} \psi[A_{\bar{z}}],
}
which is solved by\cite{Polyakov1983121,Polyakov1984223}
\al{
\label{eq:cswf}
\psi[A_{\bar{z}}]=exp\big( kS_{WZW}(U)\big).
}

\subsubsection{The Gauge Invariant Measure}

The metric of  the space of gauge potentials ($\mathscr{A}$) is given by \cite{Karabali1996135}
\al{
\bs
ds^2_{\mathscr{A}}=&\int d^2x\ \delta A^a_i \delta A^a_i=-8\int Tr(\delta A_{\bar{z}} \delta A_z)\\
=& 8 \int Tr[D_{\bar{z}}(\delta U U^{-1})D_z(U^{\dagger -1}\delta U^{\dagger})].
\es
}
This metric is similar to the Cartan-Killing metric for $SL(N,\mathbb{C})$, which is given by
\al{
ds^2_{SL(N,\mathbb{C})}=8\int Tr[(\delta U U^{-1})(U^{\dagger -1}\delta U^{\dagger})].
}
The volumes of these two spaces are related by
\al{
d\mu(\mathscr{A})=det(D_{\bar{z}}D_z)d\mu(U,U^{\dagger}).
}
To make this measure gauge invariant, we need to define a new gauge invariant matrix $H=U^{\dagger}U$, which is an element of the coset $SL(N,\mathbb{C})/SU(N)$. To integrate over only the gauge invariant combinations of $U$ and $U^\dagger$, we write
\al{
\label{eq:CSmeasure}
d\mu(\mathscr{A})=det(D_{\bar{z}}D_z)d\mu(H).
}
The determinant is given by\cite{Bos:1989kn,Nair:2005iw}
\al{
det(D_{\bar{z}}D_z)=constant \times e^{2c_AS_{WZW}(H)}
}
where $c_A$ is the quadratic Casimir in the adjoint representation.

Now, using the Polyakov-Wiegmann(PW)\cite{Polyakov1983121,Polyakov1984223} identity, we can finally write the inner product
\al{
\label{eq:cspsipsi}
\langle \psi|\psi \rangle =\int d\mu(\mathscr{A})\ e^{-K}\ \psi^*\psi=\int d\mu(H)\ e^{(2c_A+k)S_{WZW}(H)}.
}

\subsubsection{Wilson Loops}

In the temporal gauge, finding the expectation value of a Wilson loop using geometric quantization is problematic. In this gauge, the operator is given by
\al{
\label{eq:W1}
W_R(C)=Tr_R\ \mc{P}\ e^{-\oint \limits_c (A_zdz+A_{\bar{z}}d\bar{z})}.
}
The derivative operator $A_z$ in the Wilson loop makes it difficult to calculate the path ordered exponential acting on the wave-functional. To go around this problem paying almost no price, we will use a Wilson loop-like observable $\mc{W}(C)=Tr\ U(x,x,C)$, where $U$ is given by \eqref{eq:U}. We can write $\mc{W}$ as
\al{
\label{eq:W2}
\mc{W}_R(C)=Tr_R\ \mc{P}\ e^{-\oint \limits_c (\mc{A}_zdz+A_{\bar{z}}d\bar{z})}.
}
$\mc{A}_z$ is defined as $\partial_z A_{\bar{z}}-\partial_{\bar{z}}\mc{A}_z+[\mc{A}_z,A_{\bar{z}}]=0$. Since Gauss' law forces $F_{z\bar{z}}=0$, we can say that $\mc{W}$ is the Wilson loop on the constraint hypersurface. 

Since the theory is given by $S_{WZW}(H)$, WZW currents $J_{\bar{z}}=-\partial_{\bar{z}}HH^{-1}$ and $J_z=H^{-1}\partial_z H$ can be used to write gauge invariant observables. $\mc{A}_z$ and $A_{\bar{z}}$ can be written as
\al{
\label{eq:AJ}
\bs
A_{\bar{z}}=&-\partial_{\bar{z}}UU^{-1}=U^{\dagger-1}J_{\bar{z}}U^{\dagger}+U^{\dagger-1}\partial_{\bar{z}}U^{\dagger},\\
\mc{A}_z=&-\partial_zUU^{-1}=U^{\dagger-1}J_zU^{\dagger}+U^{\dagger-1}\partial_zU^{\dagger},
\es
}
which are $SL(N,\mathbb{C})$ transformed WZW currents. With this information, we can write \eqref{eq:W2} in terms of $H$ as
\al{
\mc{W}_R(C,H)=Tr_R\ \mc{P}\ e^{\ \oint \limits_c (\partial_zHH^{-1}dz+\partial_{\bar{z}}HH^{-1}d\bar{z})}.
}
Now that $\mc{W}$ is written in terms of $H$ and it commutes with the wave-functional, we can write its expectation value as 
\al{
\langle \mc{W}(C) \rangle = \int d\mu(H)\ e^{(2c_A+k)S_{WZW}(H)} \mc{W}(C,H).
}


\section{Topologically Massive Yang-Mills Theory}

The TMYM action is given by
\al{
\label{eq:actiontmym}
\bs
S_{TMYM}=&S_{CS}+S_{YM}\\
=&-\frac{k}{4\pi}\int \limits_{\Sigma \times[t_i,t_f]} {d^3x }\ \epsilon^{\mu\nu\alpha}\ Tr \pa{A_\mu \partial_{\nu} A_{\alpha} + \frac{2}{3}A_\mu A_\nu A_\alpha}\\
&-\frac{k}{4\pi}\frac{1}{4m}\int \limits_{\Sigma\times[t_i,t_f]} {d^3x }\ Tr\  F_{\mu\nu}F^{\mu\nu}.
\es
} 
By defining
\al{
\label{eq:tilde0}
\tilde A_{\mu}=A_{\mu}+\frac{1}{2m}\epsilon_{\mu\alpha\beta}F^{\alpha\beta},
}
the conjugate momenta are given by 
\al{
\Pi^{a z}=\frac{ik}{4\pi} \tilde A^a_{\bar{z}}~\ \text{and}~\ \Pi^{a \bar{z}}=-\frac{ik}{4\pi} \tilde A^a_z.
}
Using $E_{z}=\frac{i}{2m}F^{0\bar{z}}$ and $E_{\bar{z}}=-\frac{i}{2m}F^{0z}$, components of $\tilde{A}$ can be written as $\tilde A_{z}=A_{z}+E_{z}$ and $\tilde A_{\bar{z}}=A_{\bar{z}}+E_{\bar{z}}$.

The symplectic two-form of the theory is given by
\al{
\label{eq:omegatmym}
\Omega=&\frac{ik}{4\pi}\int \limits_{\Sigma}(\delta \tilde A^a_{\bar{z}} \delta A^a_z+\delta A^a_{\bar{z}}\delta \tilde A^a_z ).
}
From \eqref{eq:omegatmym}, it can be seen that the phase space of TMYM theory consists of two CS-like halves. This becomes more clear in coordinates $B_z=\frac{1}{2}(A_1+i\tilde A_2)$, $C_z=\frac{1}{2}(\tilde A_1+i A_2)$. In these coordinates the sympectic two-form is in the form of $\delta B_z \delta B_{\bar{z}} + \delta C_z \delta C_{\bar{z}}$.


\subsection{The Wave-Functional}

Choosing the holomorphic polarization leads to $\Phi[A_z,A_{\bar{z}},\tilde A_z, \tilde A_{\bar{z}}]=e^{-\frac{1}{2}K}\psi[A_{\bar{z}},\tilde A_{\bar{z}}]$, where $K=\frac{k}{4\pi}\int_\Sigma (\tilde A^a_{\bar{z}} A^a_z+ A^a_{\bar{z}} \tilde A^a_z )$ is the K\"ahler potential. Upon quantization, we write
\al{
\label{eq:deltatmym}
A^a_z\psi=\frac{4\pi}{k} \frac{\delta}{\delta \tilde A^a_{\bar{z}}} \psi\ \ ~\text{and}\ \ ~\tilde A^a_z\psi=\frac{4\pi}{k} \frac{\delta}{\delta A^a_{\bar{z}}}\psi.
}

\subsubsection{The Gauss Law}

An infinitesimal gauge transformation on $\psi[A_{\bar{z}},\tilde A_{\bar{z}}]$ is given by
\al{
\label{eq:infgauge}
\delta_\epsilon \psi[A_{\bar{z}},\tilde{A}_{\bar{z}}]=\int d^2z\ \pa{ \delta_\epsilon A^a_{\bar{z}} \frac{\delta\psi}{\delta A^a_{\bar{z}}}  +\ \delta_\epsilon \tilde{A}^a_{\bar{z}} \frac{\delta\psi}{\delta \tilde{A}^a_{\bar{z}}}}.
}
Using \eqr{eq:deltatmym} and $\delta A^a_{\bar{z}}=D_{\bar{z}}\epsilon^a$, $\delta \tilde{A}^a_{\bar{z}}=\tilde{D}_{\bar{z}}\epsilon^a$, we get
\al{
\bs
\delta_\epsilon \psi=& \int d^2z\ \epsilon^a \pa{\tilde{D}_{\bar{z}}\frac{\delta}{\delta \tilde{A}^a_{\bar{z}}} + D_{\bar{z}}\frac{\delta}{\delta A^a_{\bar{z}}} } \psi\\
=&\frac{k}{4\pi} \int d^2z\ \epsilon^a \pa { \partial_z \tilde{A}^a_{\bar{z}}+ \partial_z A^a_{\bar{z}} -2F_{z\bar{z}}- D_z E_{\bar{z}}+D_{\bar{z}}E_z  }\psi
\es
}
For TMYM theory, Gauss' law is given by $G^a\psi=0$, where $G^a=\frac{ik}{4\pi}(2F_{z\bar{z}} + D_z E_{\bar{z}} - D_{\bar{z}}E_z)$. After applying Gauss' law, the gauge transformation becomes
\al{
\label{eq:infg2}
\delta_\epsilon \psi= \frac{k}{4\pi} \int d^2z\ \epsilon^a \pa{\partial_z \tilde A^a_{\bar{z}}+\partial_z A^a_{\bar{z}} } \psi.
}
This condition is very similar to \eqref{eq:infg} and can be solved by $\psi=\phi\chi$, with
\al{
\label{eq:wftmym}
\phi[A_{\bar{z}},\tilde{A}_{\bar{z}}]=exp\pb{\frac{k}{2}\big(S_{WZW}(\tilde U)+S_{WZW}( U)\big)}.
}
Here we used the parametrization 
\al{
\label{eq:Utilde}
\tilde{U}(x,0,C)=\mc{P}exp\pa{-\underset{C}{\ \ \int_0^x}(\tilde{A}_{\bar{z}}d\bar{z}+\tilde{\mc{A}}_zdz)},
}
which can be followed by the tilde versions of the equations \eqref{eq:para} and \eqref{eq:scriptA}.

\subsubsection{Schr\"odinger's Equation}

With $\alpha=\frac{4\pi}{k}$, $B=\frac{ik}{2\pi}F^{z\bar{z}}$ and using Euclidean metric, the Hamiltonian for TMYM is given by
\al{
\mc{H}=\frac{m}{2\alpha}(E^a_{\bar{z}} E^a_z + E^a_z E^a_{\bar{z}})+\frac{\alpha}{m} B^a B^a.
}
$E$-fields satisfy the commutator
\al{
\label{eq:ecom}
[E^a_z(x),E^b_{\bar{z}}(x')]=-2 \alpha\ \delta^{ab}\delta^{(2)}(x-x').
}
This allows us to interpret $E^a_z$ as an annihilation operator and $E^b_{\bar{z}}$ as a creation operator \cite{Grignani1997360}. To get rid of the infinity, the Hamiltonian can be normal ordered as 
\al{
\mc{H}=\frac{m}{\alpha}E^a_{\bar{z}} E^a_z +\frac{\alpha}{m} B^a B^a.
}
At large distances compared to $1/m$, it is the standard practice to neglect the $B^2$ term\cite{Grignani1997360}. Since we are interested in the \emph{near} CS limit of the theory, we take $m$ to be large and ignore the potential energy term. In that case, the vacuum wave-functional is given by $E_z\psi=0$. In terms of the derivative operator, $E_z=\tilde A_z-A_z$ can be written as
\al{
E^a_z\psi=&\frac{4\pi}{k}\pa{\frac{\delta \phi}{\delta A^a_{\bar{z}}}-\frac{\delta \phi}{\delta \tilde{A}^a_{\bar{z}}}}\chi+\frac{4\pi}{k}\pa{\frac{\delta \chi}{\delta A^a_{\bar{z}}}-\frac{\delta \chi}{\delta \tilde{A}^a_{\bar{z}}}}\phi.
}
The derivatives of $\phi$ are given by $\mc{A}$ and $\tilde{\mc{A}}$\cite{Nair:2005iw} as
\al{
\label{eq:phi}
\tilde A^a_z\phi=\frac{4\pi}{k}\frac{\delta \phi}{\delta A^a_{\bar{z}}}=\mc{A}^a_z \phi\  ~\text{and}~\  A^a_z\phi=\frac{4\pi}{k}\frac{\delta \phi}{\delta \tilde{A}^a_{\bar{z}}}=\tilde{\mc{A}}^a_z \phi.
}
Now by defining $\mc{E}_z=\tilde{\mc{A}}_z - \mc{A}_z$ we can write
\al{
\label{eq:EE}
E^a_z\psi=-\mc{E}^a_z\psi+\frac{4\pi}{k}\pa{\frac{\delta \chi}{\delta A^a_{\bar{z}}}-\frac{\delta \chi}{\delta \tilde{A}^a_{\bar{z}}}}\phi.
}
Solving $E_z\psi=0$ for $\chi$ gives
\al{
\label{eq:chi}
\bs
\chi^{ }=&exp \pa{-\frac{k}{8\pi}\int \limits_{\Sigma} (\tilde{A}^a_{\bar{z}}-A^a_{\bar{z}})\mc{E}^a_z}=exp \pa{-\frac{k}{8\pi}\int \limits_{\Sigma}E^a_{\bar{z}}\mc{E}^a_z}.
\es
}
Since both $E$ and $\mc{E}$ are first order in $1/m$, $\chi=1+\mc{O}(1/m^2)$. Thus $\chi$ can be taken as unity at large enough distances compared to $1/m$.


\subsubsection{The Gauge Invariant Measure}

For TMYM theory, the metric of  the space of gauge potentials is given by
\al{
\bs
ds^2_{\mathscr{A}}=&-4\int Tr(\delta \tilde{A}_{\bar{z}} \delta A_z+\delta A_{\bar{z}}\delta \tilde{A}_z)\\
=&\ 4 \int Tr[\tilde{D}_{\bar{z}}(\delta \tilde{U} \tilde{U}^{-1})D_z(U^{\dagger -1}\delta U^{\dagger})+D_{\bar{z}}(\delta U U^{-1})\tilde{D}_z(\tilde{U}^{\dagger -1}\delta \tilde{U}^{\dagger})].
\es
}
The gauge invariant measure for this case is
\al{
\label{eq:meas}
d\mu(\mathscr{A})=det(\tilde{D}_{\bar{z}}D_z) det(D_{\bar{z}}\tilde{D}_z)d\mu(\tilde{U}^{\dagger}U)d\mu(U^{\dagger}\tilde{U}).
}
For a certain choice of local counter terms, the determinants can be written as
\al{
det(\tilde{D}_{\bar{z}}D_z) det(D_{\bar{z}}\tilde{D}_z)=constant \times e^{2c_A\big(S_{WZW}(\tilde{U}^{\dagger}U)+S_{WZW}(U^{\dagger}\tilde{U})\big)}.
}
To simplify the notation we define a new gauge invariant matrix $N=\tilde{U}^{\dagger}U$. Now, the measure becomes
\al{
\label{eq:measure}
d\mu(\mathscr{A})=constant \times e^{2c_A\big(S_{WZW}(N)+S_{WZW}(N^{\dagger})\big)} d\mu(N)d\mu(N^{\dagger}).
}

\subsubsection{Chern-Simons Splitting}

To find the inner product, using PW identity we write
\al{
e^{-K_{TMYM}}\psi_{TMYM}^*\psi^{ }_{TMYM}=e^{\frac{k}{2}\big(S_{WZW}(N)+S_{WZW}(N^{\dagger})\big)}\chi^*\chi.
}
We have shown that $\chi^*\chi=1+\mc{O}(1/m^2)$.
Then the inner product for the vacuum state in the near CS limit is
\al{
\label{eq:psipsi}
\langle \psi|\psi\rangle_{TMYM}= \int d\mu(N)d\mu(N^{\dagger})\ e^{(2c_A+\frac{k}{2})\big(S_{WZW}(N)+S_{WZW}(N^{\dagger})\big)}+\mc{O}(1/m^2).
}
Using \eqref{eq:cspsipsi}, this inner product can be written as two CS theories with half the level, as
\al{
\label{eq:cssplitting}
\langle \psi|\psi\rangle_{TMYM_k}=\langle\psi|\psi\rangle_{CS_{k/2}}\langle\psi|\psi\rangle_{CS_{k/2}}+\mc{O}(1/m^2).
}
Here the labels $k$ and $k/2$ indicate the levels of the CS terms in the Lagrangian.
It is well known that $k$ has to be an integer to ensure gauge invariance of CS. Here, we have two half level CS parts and each piece transforms as $\frac{1}{2}S_{CS}(A^g)\rightarrow\frac{1}{2}S_{CS}(A)+\pi k \omega(g)$ where $\omega(g)$ is the winding number. Then, the sum of the two will bring an extra $2\pi k \omega(g)$ that will not change the value of the path integral, even for odd values of $k$. But if one wants to write operator expectation values of TMYM theory in terms of CS expectation values using CS splitting, this can only be done for even values of $k$.


\section{Wilson Loops in TMYM Theory}\label{sec:W}

Since $\tilde A$ transforms like a gauge field, we can define a Wilson loop-like observable with it, as
\al{
\label{eq:tildewlsn}
T_R(C)=Tr_R\ \mc{P}\ e^{-\oint \limits_c \tilde{A}_\mu dx^\mu}.
}
To make a physical interpretation of this new observable, we will check and see if it satisfies a 't Hooft-like algebra with the Wilson loop. To simplify the calculation we will consider the abelian versions of the loop operators. In complex coordinates with temporal gauge, the operators are
\al{
W(C)=e^{i\oint \limits_c (A_z dz+A_{\bar{z}} d\bar{z})}\ ~ \text{and}\ ~ T(C)=e^{i\oint \limits_c (\tilde{A}_z dz+\tilde{A}_{\bar{z}} d\bar{z})}.
}
These operators satisfy the following 't Hooft-like algebra
\al{
\label{eq:thooft}
T(C_1)W(C_2)=e^{\frac{2\pi i}{k}l(C_1,C_2)}W(C_2)T(C_1),
}
where $l(C_1,C_2)$ is the intersection number of $C_1$ and $C_2$ , which can only take values $0,\pm 1$. This allows us to interpret $T$ as a 't Hooft-like operator for TMYM theory. In this work, we will only consider loops that have zero intersection number.

Instead of directly using $T$ and $W$, we will use $Tr\ U(x,x,C)$ and $Tr\ \tilde{U}(x,x,C)$ or
\al{
\mc{W}_R(C)=Tr_R\ \mc{P}\ e^{-\oint \limits_c (\mc{A}_zdz+A_{\bar{z}}d\bar{z})}\ ~
\text{and}~ \ 
\mc{T}_R(C)=Tr_R\ \mc{P}\ e^{-\oint \limits_c (\tilde{\mc{A}}_zdz+\tilde{A}_{\bar{z}}d\bar{z})}
}
to avoid the same problem we had in CS Wilson loops. Once again these can be written using gauge $SL(N,\mathbb{C})$ transformed WZW currents $-\partial_{\bar{z}}NN^{-1}$,  $-\partial_zNN^{-1}$,  $-\partial_{\bar{z}}N^{\dagger}N^{\dagger-1}$ and $-\partial_zN^{\dagger}N^{\dagger-1}$ as
\al{
\bs
&\mc{W}_R(C,N)=Tr_R\ \mc{P}\ e^{\ \oint \limits_c (\partial_zNN^{-1}dz+\partial_{\bar{z}}NN^{-1}d\bar{z})},\\
&\mc{T}_R(C,N^\dagger)=Tr_R\ \mc{P}\ e^{\ \oint \limits_c (\partial_zN^{\dagger}N^{\dagger-1}dz+\partial_{\bar{z}}N^{\dagger}N^{\dagger-1}d\bar{z})}.
\es
}
Using these operators, we can calculate the following expectation value
\al{
\bs
\langle \mc{W}_{R_1}(C_1)&\mc{T}_{R_2}(C_2) \rangle=\int d\mu(\mathscr{A}) \psi_0^*\mc{W}_{R_1}(C_1)\mc{T}_{R_2}(C_2)\psi_0\\
=&\int d\mu(N)d\mu(N^{\dagger})\ e^{(2c_A+\frac{k}{2})\big(S_{WZW}(N)+S_{WZW}(N^{\dagger})\big)}\ \mc{W}_{R_1}(C_1,N)\mc{T}_{R_2}(C_2,N^{\dagger})\\
&+\mc{O}(1/m^2).
\es
}
This leads to interesting equivalences between the observables of TMYM and CS theories:
\begin{subequations}
\label{eq:equivalence}
\al{
\label{eq:equivalence1}
\langle \mc{W}_R(C)\rangle_{TMYM_{2k}} = \langle \mc{W}_R(C)\rangle_{CS_{k}}+\mc{O}(1/m^2),
}
\al{
\label{eq:equivalence2}
\langle \mc{T}_R(C)\rangle_{TMYM_{2k}} = \langle \mc{W}_R(C)\rangle_{CS_{k}}+\mc{O}(1/m^2)
}
and
\al{
\label{eq:equivalence3}
\langle \mc{W}_{R_1}(C_1)\mc{T}_{R_2}(C_2)\rangle_{TMYM_{2k}} = \bigg(\langle \mc{W}_{R_1}(C_1)\rangle_{CS_{k}}\bigg)\bigg(\langle \mc{W}_{R_2}(C_2)\rangle_{CS_{k}}\bigg)+\mc{O}(1/m^2).
}
\end{subequations}
Notice that these equivalences are written for even level number on the TMYM side. Although $\langle \mc{W}_R(C_1)\mc{T}_R(C_2)\rangle_{TMYM_{k}}$ is gauge invariant for all integer values of $k$, \eqref{eq:equivalence} can only be written for even level number on the TMYM side.


\section{Pure Yang-Mills Theory}

In previous sections, we have shown that at large distances, TMYM theory exhibits a CS+CS type splitting behavior, analogous to \eqref{eq:tmgcs+cssplitting}. This section is a quick summary of how CS-CS splitting can be obtained for pure YM theory at large distances, analogous to \eqref{eq:YMlimit}.

The action is given by
\al{
\label{eq:actionym}
S_{YM}=-\frac{k}{4\pi}\frac{1}{4m}\int \limits_{\Sigma\times[t_i,t_f]} {d^3x }\ Tr\  (F_{\mu\nu}F^{\mu\nu}).
}
Here, the constant $\frac{k}{4\pi}$ is inserted so that the split CS terms will have level numbers at the end of our calculation. At this point there are no restrictions on $k$.
The symplectic two-form of the theory is
\al{
\label{eq:omegaym}
\Omega=\frac{ik}{4\pi}\int \limits_{\Sigma}(\delta E^a_{\bar{z}} \delta A^a_z+\delta A^a_{\bar{z}}\delta E^a_z).
}
With defining
\al{
\label{eq:tildehatdef}
\tilde A_i=A_i+E_i\ ~\text{and}~\ \hat A_i=A_i-E_i,
}
the symplectic two-form can be written as a difference of two half CS-like parts as
\al{
\label{eq:omegaym2}
\Omega=\frac{ik}{4\pi}\int \limits_{\Sigma}(\delta \tilde A^a_{\bar{z}} \delta A^a_z-\delta A^a_{\bar{z}}\delta \hat A^a_z).
}
Using the methods we described in the previous sections, it can be shown that\cite{yildirim2} the wave-functional is given by
\al{
\label{eq:wfym1}
\psi[A_{\bar{z}},\tilde{A}_{\bar{z}}]=exp\pb{\frac{k}{2}\big(S_{WZW}(\tilde U)-S_{WZW}( U)\big)}\chi,
}
or equally
\al{
\label{eq:wfym2}
\psi[A_{\bar{z}},\hat{A}_{\bar{z}}]=exp\pb{\frac{k}{2}\big(S_{WZW}(U)-S_{WZW}( \hat U)\big)}\chi.
}
Similar to TMYM theory, it can be shown that $\chi=1+\mc{O}(1/m^2)$, since both TMYM theory and pure YM theory have the same Hamiltonian in the temporal gauge. 

By defining new gauge invariant matrices $H_1=U^{\dagger}\tilde{U}$ and $H_2=\hat{U}^{\dagger}U$, the measure is given by
\al{
\label{eq:meas2}
d\mu(\mathscr{A})=e^{2c_A\big(S_{WZW}(H_1)+S_{WZW}(H_2)\big)}d\mu(H_1)d\mu(H_2).
}
Now, using PW identity, the inner product can be written as
\al{
\label{eq:YMinnerproduct}
\langle \psi|\psi \rangle=\int d\mu(H_1)d\mu(H_2)e^{\pa{2c_A+\frac{k}{2}}S_{WZW}(H_1)+\pa{2c_A-\frac{k}{2}}S_{WZW}(H_2)}+\mc{O}(1/m^2).
}
This inner product can also be written in terms of two CS inner products with levels $k/2$ and $-k/2$ as
\al{
\label{eq:YMsplit}
\langle \psi_0|\psi_0\rangle_{YM_k}=\langle \psi|\psi \rangle_{CS_{k/2}}\langle \psi|\psi \rangle_{CS_{-k/2}}+\mc{O}(1/m^2).
}
This shows that the insight provided by the gravitational analogue theory was correct in this limit as well. Analogous to the pure Hilbert-Einstein limit \eqref{eq:YMlimit}, pure YM theory can also be written as a difference of two split half-CS parts at large finite distances. 

Although initially there were no restrictions on $k$, now it may seem like it has to be an even integer to ensure gauge invariance of the split CS terms. This is not necessary, since each CS part transforms like $\frac{1}{2}S_{CS}(A^g)\rightarrow\frac{1}{2}S_{CS}(A)+\pi k \omega(g)$ and two $\pi k \omega(g)$ terms cancel each other. But if one wants to write YM observables in terms of CS Wilson loops, similar to our discussion on TMYM observables, this can only be done for even values of $k$.


\section{Conclusions}

We have shown that at large enough distances, both near CS limit and pure YM limit of TMYM theory exhibits a CS splitting behavior as predicted by the analogous gravitational theory, topologically massive AdS gravity. In the near CS limit, each split CS part has the level $k/2$. The pure YM theory however, has split parts with levels $k/2$ and $-k/2$. At very large distances, split CS parts of TMYM theory add up to give the original level number $k$. On the other hand, for pure YM theory, the split parts cancel at very large distances to give a trivial result, as required by the existence of a mass gap. 

In section \ref{sec:W}, we have shown that the CS splitting can be exploited to write TMYM observables in terms of CS observables. This allows us to use skein relations in order to calculate TMYM Wilson loop expectation values. A similar calculation can be done for YM observables at large enough distances.

Detailed calculations on this work can be found in refs. \citen{yildirim, yildirim2, yildirimtez}.


\section*{Acknowledgements}

I thank Vincent Rodgers and Parameswaran Nair for their support and supervision. I also would like to thank Tudor Dimofte and Chris Baesley for helpful discussions, and conference organizers for the support they provided.


\bibliography{Bibliography.bib}

\begin{thebibliography}{10}

\bibitem{Deser1982372}
S.~Deser, R.~Jackiw, and S.~Templeton.
\newblock {Topologically Massive Gauge Theories}.
\newblock {\em Annals of Physics}, 140:372 -- 411, 1982.

\bibitem{Deser2}
S.~Deser, R.~Jackiw, and S.~Templeton.
\newblock {Three-Dimensional Massive Gauge Theories}.
\newblock {\em Physical Review Letters}, 48:975--978, 1982.

\bibitem{achucarro1986chern}
A.~Ach{\'u}carro and P.K. Townsend.
\newblock {A Chern-Simons action for three-dimensional anti-de Sitter
  supergravity theories}.
\newblock {\em Physics Letters B}, 180(1):89--92, 1986.

\bibitem{witten1988}
E.~Witten.
\newblock {2+1 Dimensional Gravity as an Exactly Soluble System}.
\newblock {\em Nuclear Physics B}, 311(1):46--78, 1988.

\bibitem{Carlip2008272}
S.~Carlip, S.~Deser, A.~Waldron, and D.K. Wise.
\newblock {Topologically Massive AdS Gravity}.
\newblock {\em Physics Letters B}, 666(3):272 -- 276, 2008.

\bibitem{Carlip2}
S.~Carlip, S.~Deser, A.~Waldron, and D.K. Wise.
\newblock {Cosmological Topologically Massive Gravitons and Photons}.
\newblock {\em Classical and Quantum Gravity}, 26(7):075008, 2009.

\bibitem{Bos:1989kn}
M.~Bos and V.P. Nair.
\newblock {Coherent State Quantization of Chern-Simons Theory}.
\newblock {\em International Journal of Modern Physics A}, A5:959, 1990.

\bibitem{Nair:2005iw}
V.P. Nair.
\newblock {\em {Quantum Field Theory: A modern Perspective}}.
\newblock Springer, 2005.

\bibitem{Karabali:1996iu}
D.~Karabali and V.P. Nair.
\newblock {Gauge Invariance and Mass Gap in (2+1)-Dimensional Yang-Mills
  Theory}.
\newblock {\em International Journal of Modern Physics A}, A12:1161--1172,
  1997.

\bibitem{Polyakov1983121}
A.~Polyakov and P.B. Wiegmann.
\newblock {Theory of Nonabelian Goldstone Bosons In Two Dimensions}.
\newblock {\em Physics Letters B}, 131:121 -- 126, 1983.

\bibitem{Polyakov1984223}
A.M. Polyakov and P.B. Wiegmann.
\newblock {Goldstone Fields In Two Dimensions With Multivalued actions}.
\newblock {\em Physics Letters B}, 141:223 -- 228, 1984.

\bibitem{Karabali1996135}
D.~Karabali and V.P. Nair.
\newblock {A Gauge-Invariant Hamiltonian Analysis for Non-Abelian Gauge
  Theories in (2+1) Dimensions}.
\newblock {\em Nuclear Physics B}, 464:135 -- 152, 1996.

\bibitem{Grignani1997360}
G.~Grignani, G.~Semenoff, P.~Sodano, and O.~Tirkkonen.
\newblock {G/G Models as the Strong Coupling Limit of Topologically Massive
  Gauge Theory}.
\newblock {\em Nuclear Physics B}, 489:360 -- 384, 1997.

\bibitem{yildirim2}
T.~Yildirim.
\newblock {Chern-Simons Splitting of 2+1D Pure Yang-Mills Theory at Large
  Distances}.
\newblock {\em arXiv preprint arXiv:1410.8593}, 2014.

\bibitem{yildirim}
T.~Yildirim.
\newblock {Topologically Massive Yang-Mills Theory and Link Invariants}.
\newblock {\em To appear in International Journal of Modern Physics A, arXiv
  preprint arXiv:1311.1853}, 2013.

\bibitem{yildirimtez}
Tuna Yildirim.
\newblock {\em Topologically Massive Yang-Mills Theory and Link Invariants}.
\newblock PhD thesis, The University of Iowa, arXiv:1412.4310, 2014.

\end{thebibliography}
\bibliographystyle{unsrt}

\end{document}